\documentclass[preprint,prd,nofootinbib,tightenlines,amsmath]{revtex4}
\usepackage{graphicx}
\usepackage{bm}
\usepackage{color}
\usepackage{subfigure}

\def\lsim{\mathrel{\rlap{\lower4pt\hbox{\hskip1pt$\sim$}}
    \raise1pt\hbox{$<$}}}         
\def\gsim{\mathrel{\rlap{\lower4pt\hbox{\hskip1pt$\sim$}}
    \raise1pt\hbox{$>$}}}         

\begin{document}
\baselineskip=15pt \parskip=5pt

\vspace*{3em}

\hfill{NPAC 11-05}

\title{R$\nu$MDM and Lepton Flavor Violation}

\author{Yi Cai$^1$, Xiao-Gang He$^{1,2}$, Michael Ramsey-Musolf$^{3,4}$ and Lu-Hsing Tsai$^2$}

\affiliation
{$^1$INPAC, Department of Physics, Shanghai Jiao Tong University, Shanghai, China\\
$^2$Department of Physics and  Center for Theoretical Sciences, \\
National Taiwan University, Taipei, Taiwan\\
$^3$ Department of Physics, University of Wisconsin-Madison, Madison, WI USA\\
$^4$ California Institute of Technology, Pasadena, CA USA}

\date{\today $\vphantom{\bigg|_{\bigg|}^|}$}

\begin{abstract}
A model relating radiative seesaw and minimal dark matter mass scales without beyond the standard model (SM) gauge symmetry (R$\nu$MDM) is constructed.   In addition to the SM particles, the R$\nu$MDM contains, a Majorana fermion multiplet $N_R$ and scalar multiplet $\chi$ that transform respectively as $(1,5,0)$  and $(1,6,-1/2)$ under the SM gauge group $SU(3)_C\times SU(2)_L\times U(1)_Y$.  The neutral component $N_R^0$ plays the role of dark matter with a mass in the range of 9 to 10 TeV.  This scale also sets the lower limit for the scale for the heavy degrees of freedom in $N_R$ and $\chi$ which generate light neutrino masses through the radiative seesaw mechanism.  The model predicts an $N_R^0$-nucleus scattering cross section that would be accessible with future dark matter direct detection searches as well as observable effects in present and searches for charged lepton flavor violating processes, such as $l_i\to l_j \gamma$ and $\mu - e$ conversion.

\end{abstract}

\maketitle

\section{Introduction}

To account for the cosmological observation that our universe is composed of about~\cite{pdg,silk} 20\% dark matter (DM), the standard model (SM) of electroweak and strong interactions has to be extended beyond its minimal form.  In particular, new particles playing the role of DM have to be introduced.  One of the most popular candidates is a weakly interacting massive particle (WIMP) that leaves a thermal relic density below the relevant freeze-out temperature in the early universe.  There are many possible ways to introduce such a WIMP candidate.  To guarantee that the WIMP is stable, most of the models employ a symmetry beyond the SM gauge symmetries, such as R-parity in SUSY theories~\cite{silk}, or a $Z_2$ symmetry in dark matter models with a real or complex singlet~\cite{singlet}.

It is interesting to ask whether the SM gauge symmetry alone can already stabilize a new particle such that it may be a WIMP candidate.  Indeed, it has been shown that such possibility can be implemented using the minimal dark matter (MDM) idea~\cite{cerilli}.  The MDM scenario achieves stabilization of a WIMP by choosing representations of the $SU(2)_L\times U(1)_Y$ gauge group such that (a) the choice of hypercharge implies the existence of an electrically neutral component of the multiplet, (b) the neutral component has the lowest mass after accounting for radiatively-induced mass splittings, and (c) they do not couple to the SM decay products directly at the renormalizable level\footnote{Higher-dimension, non-renormalizable operators that would generate decays can be suppressed if the associated mass scale is sufficiently large. }.  Therefore, the lightest component field does not decay into SM particles, making it a DM candidate.  Such a model has an additional bonus that the DM mass is completely fixed by the observed DM relic density because the DM thermal relic density is produced by known electroweak interactions.

Apart from the dark matter problem, the observation of neutrino oscillations~\cite{pdg} necessitates extending the minimal SM.
The simplest extension introduces gauge-singlet right-handed neutrinos as needed to construct Dirac mass operators.  However the vast hierarchy between the tiny Yukawa couplings associated with the Dirac mass operators and those associated with the charged lepton masses is widely considered theoretically unattractive, and many models, including  seesaw models~\cite{seesaw1,seesaw2,seesaw3} and the Zee model~\cite{zee}, have been developed seeking a more \lq\lq natural" explanation of the neutrino mass scale.  Common to these approaches is the existence of new particles with large masses above electroweak scale.  We refer these large mass scales collectively as the neutrino mass scale, which is not determined {\em a priori}.

It has been shown in Ref.\cite{e-ma} that it is possible to link the DM mass scale to the neutrino mass scale by generating neutrino masses radiatively, leading to the radiative seesaw model (RSSM).  This idea has been further studied by various authors~\cite{other-neu}.  Most of the models proposed impose discrete symmetries beyond the SM gauge symmetry to stabilize DM. In this work, we explore the possibility of relating DM and neutrino mass scales without imposing additional discrete symmetries on the originally renormalizable Lagrangian by combining the MDM and RSSM ideas.  We refer to this model as  the \lq\lq R$\nu$MDM" scenario.  In particular, we demonstrate that it is indeed possible to construct consistent theoretical models to achieve this goal and that they may have testable phenomenological predictions.
In the following we describe a minimal model of this type and study some of its implications for charged lepton flavor violation (LFV). We show that while this version of the R$\nu$MDM is unlikely to be testable with the Large Hadron Collider, it can lead to observable effects in future dark matter direct detection experiments and searches for LFV.

\section{The model}

As we explain below, the minimal model of this type contains a new right-handed Majorana fermion $N_R$ and  scalar $\chi$ multiplet that transform under the SM gauge group $SU(3)_C\times SU(2)_L\times U(1)_Y$ as
\begin{eqnarray}
N_R: (1, 5, 0)\;,\;\;\chi: (1, 6, -1/2)\;.
\end{eqnarray}
In addition to the terms already existing in the minimal SM, the most general renormalizable interactions containing the $N_R$ and $\chi$ are
\begin{eqnarray}
&&L_{new} =  \bar N_R i \gamma^\mu D_\mu N_R+(D^\mu \chi)^\dagger D_\mu \chi-(\bar L_L Y \chi N_R + {1\over 2} \bar N_R^c M N_R + h.c.) - V_{new}\nonumber\\
&&V_{new} = \mu^2_\chi \chi^\dagger \chi + \lambda_\chi^\alpha (\chi \chi^\dagger \chi \chi^\dagger)_\alpha +
\lambda^\alpha_{H\chi} (H^\dagger H \chi^\dagger \chi)_\alpha + \left[ \tilde \lambda_{H\chi}(H \chi)^2 + h.c.\right]\;,
\end{eqnarray}
where $D_\mu$ is the gauge covariant derivative and $H$ is the SM Higgs doublet transforming as $(1,2,1/2)$.  Here, we include three families of $N_R$ but have suppressed the generation indices for notational simplicity. The meaning of the $\alpha$ index will be explained later.  In the present context, we do not consider possible effects of CP-violation, so will take the new couplings to be real. We defer a consideration of possible CP-violation to future work.


If $\mu^2_\chi$ and the quartic couplings in $V_{new}$ are positive, $\chi$ will not develop a non-zero vacuum expectation value (vev). It is in principle possible that renormalization group running could lead to instabilities ($\langle \chi\rangle \not=0$) at high scales, since $\chi$ couples to fermions in the third term of $L_{new}$ and since fermion loops can cause the quartic scalar couplings to run negative at large field values (see Refs.~\cite{Sher:1988mj,Casas:1994qy,Hambye:1996wb,Gonderinger:2010yn} and references therein).  However, a suitable choice of the cut-off of the theory and tree-level couplings $\lambda_\chi^\alpha$ and $\lambda^\alpha_{H\chi}$ can preclude instability of the $\langle \chi \rangle =0$ vacuum. The last operator in $V_{new}$, in contrast, receives no such potentially destabilizing contributions at one-loop, so the input value of $\tilde \lambda_{H\chi}$ can be chosen at will.  As we discuss below, this freedom is critical to the viability of the radiative seesaw mechanism.

The whole Lagrangian is invariant under a $Z_2$ symmetry even after spontaneous symmetry breaking ($\langle H\rangle \not=0$), under which $\chi \to - \chi$, $N_R \to - N_R$, and all the SM fields do not change sign.  Note that this symmetry is not imposed but rather emerges as a consequence of gauge invariance given the field content.  This accidental $Z_2$ symmetry makes the lightest particle of
the component fields in $N_R$ and $\chi$ stable, thereby providing  a WIMP candidate in each multiplet.  In addition, the active neutrinos do not have masses at tree-level.  They arise, rather, at one-loop order when integrating out the $N_R$ and $\chi$ particles.

The choice of the field content is arrived by the following considerations.  The field $N_R$ should have zero hypercharge in order to possess a Majorana mass term.  One should not allow a $\bar L H N_R$ type of Yukawa coupling in order to forbid $N_R$ from mixing with the left-handed lepton doublet and thereby allowing the $N_R^0$ to be stable.  This prevents $N_R$ from being a singlet or triplet under $SU(2)_L$.  The next possibilities are $(2,0)$ or $(4,0)$, where we now drop the SU(3$)_C$ quantum numbers.  However, the component fields in both cases are all fractionally charged ($\pm1/2$ and $\pm 3/2$), thereby precluding the possibility of a DM candidate.  In addition, if the $(2,0)$ representation is chosen for $N_R$, the accompanying Higgs representation $\chi$ needed for the Yukawa coupling term $\bar L_L N_R \chi$ required for radiative neutrino mass generation would  be required to be $(1,-1/2)$ or $(3,-1/2)$.  If $(4,0)$ is chosen for $N_R$ instead, $\chi$ can be $(3,-1/2)$ and $(5,-1/2)$.  Again the component fields in $\chi$  are all fractionally charged.  Thus if the lightest particle is in $\chi$, it will be stable with fractional charge.  These models are not phenomenologically acceptable as dark matter models.  The minimal choice for $N_R$, therefore, is $(5,0)$.  Its neutral component may be the candidate for DM.  Once the quantum numbers of the $N_R$ are fixed, the choice for $\chi$ can be made.  In order to have the heavy neutrino Yukawa term $\bar L_L \chi N_R$, $\chi$ must be either $(4, -1/2)$ or $(6,-1/2)$.  If one chooses $(4,-1/2)$, a $H^\dagger H H  \chi$ term is then allowed in the Higgs potential.  This should be forbidden because this term will induce vev for $\chi$ leading to the breaking of the accidental $Z_2$ discrete symmetry after $H$ develops vev and will also induce $\chi$ or $N_R$ to decay so that none of them are stable.

The fields $N_R$ and $\chi$ can be written in the tensor forms, $N_R = N_{ijkl}$ and $\chi = \chi_{ijklm}$.  Here the sub-indices take the values 1 and 2, and the fields are totally symmetric under exchange the sub-indices.  The independent fields $(N^{++}, N^+, N^0, N^-, N^{--})$ in $N_R$, and the independent fields $(\chi^{++}, \chi^+, \chi^0, \chi^-,\chi^{--}, \chi^{---})$ in $\chi$
can be expressed as
\begin{eqnarray}
&&N_{1111} = N^{++}\;,\;\;N_{1112} = {1\over \sqrt{4}}N^+\;,\;\;N_{1122} = {1\over \sqrt{6}} N^0\;,\nonumber\\
&&N_{1222} = {1\over \sqrt{4}}N^-\;,\;\;N_{2222} = N^{--}\;,\nonumber\\
&&\chi_{11111} = \chi^{++}\;,\;\;\chi_{11112} = {1\over \sqrt{5}}\chi^+\;,\;\;\chi_{11122}= {1\over \sqrt{10}} \chi^0\;,\nonumber\\
&&\chi_{11222} = {1\over \sqrt{10}}\chi^-\;,\;\;\chi_{12222} = {1\over \sqrt{5}}\chi^{--}\;,\;\;\chi_{22222} = \chi^{---}\;.
\end{eqnarray}

Writing the Yukawa coupling, the Majorana mass and Higgs potential terms in the tensor notation, we have
\begin{eqnarray}
&&\bar L_L \chi N_R = \bar L_i N_{jklm}\chi_{ij'k'l'm'} \epsilon^{jj'}\epsilon^{k k'}\epsilon^{l l'}\epsilon^{m m'}\;,\;\;\bar N^c_R N_R = \bar N^c_{ijkl}N_{i'j'k'l'}
\epsilon^{ii'}\epsilon^{jj'}\epsilon^{kk'}\epsilon^{ll'}\;,\nonumber\\
&&\chi\chi^\dagger = \chi_{ijklm}\chi^\dagger_{ijklm}\;,\;\;(H^\dagger H\chi^\dagger \chi)_1 = H^\dagger_a \chi_{ijklm} H_a\chi^\dagger_{ijklm}\;,\;\;(H^\dagger H\chi^\dagger \chi)_2 = H^\dagger_i \chi_{ijklm} H_a\chi^\dagger_{ajklm}\;,\nonumber\\
&&H_{a'}\chi^\dagger_{i'jklm}\epsilon^{ii'}\epsilon^{a'a}\;,
\;\;(H\chi)^2 =H_i \chi_{jklmn} H_{i'}\chi_{j'k'l'm'n'} \epsilon^{ij}\epsilon^{i'j'}\epsilon^{kk'}\epsilon^{ll'}\epsilon^{mm'}\epsilon^{nn'}\;,
\end{eqnarray}
where $\epsilon^{12} = 1$, $\epsilon^{21} = -1$
and all other elements equal to zero. In the above repeated indices are contracted by $g^{ij}$ with $g^{11} = g^{22} = 1$ and $g^{12} = g^{21} = 0$.

There are 9 different ways to write $(\chi\chi\chi^\dagger \chi^\dagger)_\alpha$ depending on how the indices are contracted by the
tensors $\epsilon^{ij}$ and $g^{ij}$. Using the identity $\epsilon^{ij}\epsilon^{kl} = g^{ik}g^{jl} - g^{il}g^{jk}$, one can show that only 3 of them are independent. The
independent terms can be chosen to be the following ones
\begin{eqnarray}
&&(\chi\chi\chi^\dagger\chi^\dagger)_1 = \chi_{ijklm}\chi^*_{ijklm}\chi_{i'j'k'l'm'}\chi^*_{i'j'k'l'm'}\;,
(\chi\chi\chi^\dagger\chi^\dagger)_2 = \chi_{ijklm}\chi^*_{ijklm'}\chi_{i'j'k'l'm'}\chi^*_{i'j'k'l'm}\;,\nonumber\\
&&(\chi\chi\chi^\dagger\chi^\dagger)_3 = \chi_{ijklm}\chi^*_{ijkl'm'}\chi_{i'j'k'l'm'}\chi^*_{i'j'k'lm}\;.
\end{eqnarray}

There are 9 different ways to write $(\chi\chi\chi^\dagger \chi^\dagger)_\alpha$ depending on how the indices are contracted by the
tensors $\epsilon^{ij}$ and $g^{ij}$. Using the identity $\epsilon^{ij}\epsilon^{kl} = g^{ik}g^{jl} - g^{il}g^{jk}$, one can show that only 3 of them are independent. The
independent terms can be chosen to be the following ones
\begin{eqnarray}
&&(\chi\chi\chi^\dagger\chi^\dagger)_1 = \chi_{ijklm}\chi^*_{ijklm}\chi_{i'j'k'l'm'}\chi^*_{i'j'k'l'm'}\;,
(\chi\chi\chi^\dagger\chi^\dagger)_2 = \chi_{ijklm}\chi^*_{ijklm'}\chi_{i'j'k'l'm'}\chi^*_{i'j'k'l'm}\;,\nonumber\\
&&(\chi\chi\chi^\dagger\chi^\dagger)_3 = \chi_{ijklm}\chi^*_{ijkl'm'}\chi_{i'j'k'l'm'}\chi^*_{i'j'k'lm}\;.
\end{eqnarray}

The Majorana mass term for $N_R$ expanded in component fields is given by
\begin{eqnarray}
\bar N^c_R M N_R &=& \bar N_R^{++c} M N_R^{--}-\bar N_R^{+c} M N_R^- + \bar N_R^{0c} M N_R^0 - \bar N_R^{-c} M N_R^+ + \bar N_R^{--c} M N_R^{++}\;\nonumber\\
&=&\bar N^{--}M P_R N^{--} + \bar N^- M P_R N^- + \bar N^0 M P_R N^0 + \bar N^+ M P_R N^+ + \bar N^{++} M P_R N^{++}\;,
\end{eqnarray}
where $P_R$ denotes the right-handed projection operator and where we have defined the four component fields $N$ in terms of $N_R$ by
\begin{eqnarray}
&&N^{--} = N^{--}_R + N^{++ c}_R\;,\quad N^- = N^-_R - N^{+ c}_R ,\quad N^0 = N^0_R + N^{0 c}_R ,\nonumber\\
&&N^{+} = N^{- c}_R - N^{+}_R \;,\quad N^{++} = N^{++}_R + N^{-- c}_R\;,
\end{eqnarray}
and where $M$ is a matrix in the space of the three $N_R$ generations.

We will work in the basis where $M = \mathrm{diag}(m_{N_1}, m_{N_2}, m_{N_3})$.  At the tree level all components of $N_j$  ($j=1,2,3$) have the same mass $m_{N_j}$.  At one-loop level, this degeneracy is lifted with a mass splitting between components of different electric charges $Q$ and $Q'$~\cite{cerilli}
\begin{eqnarray}
&&m^Q_N - m^{Q'}_N = {g^2 m_N\over 16 \pi^2}(Q^2-Q^{'2})\left [\sin^2\theta_W f(m_Z/m_N) + f(m_W/m_N) - f(m_Z/m_N)\right ]\;,\nonumber\\
&&f(x) = {x\over 2}\left\{2x^3\ln x -2 x +(x^2-4)^{1/2}(x^2+2)\ln[(x^2-2-x\sqrt{x^2-4})/2]\right\}\;.
\end{eqnarray}
For a given generation, $N^0$ has the smallest mass, making it a dark matter candidate.  The $Q=\pm$ partners of $Q=0$ component are heavier by about 166 MeV for $m_N \gg m_Z$ making the decay $N^{\pm} \to N^0 \pi^\pm$ possible.

After electroweak symmetry breaking in which only the neutral component of $H$ acquires a vev, the mass terms for the component fields of $\chi$ are given by
\begin{eqnarray}
&&m^2_{\chi^{---}} = \mu^2_\chi + \lambda_{H\chi}^1 v^2 + \lambda_{H\chi}^2 v^2\;,\nonumber\\
&&m^2_{Re\chi^{0}} = \mu^2_\chi + \lambda_{H\chi}^1 v^2 + {2\over 5} \lambda^2_{H\chi} v^2  + {6\over 5} \tilde \lambda_{H\chi} v^2\;,\nonumber\\
&&m^2_{Im\chi^{0}} = \mu^2_\chi + \lambda_{H\chi}^1 v^2 + {2\over 5} \lambda^2_{H\chi} v^2  - {6\over 5} \tilde \lambda_{H\chi} v^2\;,
\label{rimass}
\end{eqnarray}
where $v = \langle H\rangle$ is the vev of the Higgs doublet.

The $\tilde \lambda_{H\chi}$ operator mixes $\chi^+$ and $(\chi^-)^\dagger$,  $\chi^{++}$ and $(\chi^{--})^\dagger$. In the basis of $(\chi^+, (\chi^-)^\dagger)$ and $(\chi^{++}, (\chi^{--})^\dagger)$, we have the following mass matrices respectively,
\begin{eqnarray}
&&M^2_{+,-} = \left (\begin{array}{cc}
\mu^2_\chi + \lambda_{H\chi}^1 v^2 + {1\over 5} \lambda^2_{H\chi} v^2  & - {4\sqrt{2}\over 5}\tilde \lambda_{H\chi} v^2\\
- {4\sqrt{2}\over 5}\tilde \lambda_{H\chi} v^2&\mu^2_\chi + \lambda_{H\chi}^1 v^2 + {3\over 5} \lambda^2_{H\chi} v^2
\end{array}
\right )\;,\nonumber\\
&&M^2_{++,--} = \left (\begin{array}{cc}
\mu^2_\chi + \lambda_{H\chi}^1 v^2  & {2\over \sqrt{5}}\tilde \lambda_{H\chi} v^2\\
{2\over \sqrt{5}}\tilde \lambda_{H\chi} v^2&\mu^2_\chi + \lambda_{H\chi}^1 v^2 + {4\over 5} \lambda^2_{H\chi} v^2
\end{array}
\right )\;.
\end{eqnarray}
One can adjust the tree level parameters to make the particles above either heavier or lighter than the component fields in $N_R$, and also to split the mass degeneracy of the component fields in $\chi$.  Similar to the case for $N_R$, loop effects also split the mass degeneracy in the $\chi$ multiplet. For our purposes, the details of the $\chi$ spectrum are not essential, as we will take $\chi$ to be heavier than $N_R$ (see below).

\section{Dark Matter }

As mentioned earlier,  the choice  $\mu^2_\chi > 0$ implies that $\chi$ cannot develop a non-zero vev.  This leads to a residual $Z_2$ symmetry, under which $N_R\to -N_R$ and $\chi \to - \chi$ and all the other fields do not change signs. As a result the lightest component field in $\chi$ or $N_R$ will be stable and therefore can play the role of DM. We emphasize that this residual $Z_2$ symmetry is a consequence of gauge invariance, renormalizability, and concavity of the scalar potential for $\chi$ and does not result from setting any otherwise allowed couplings to zero by hand.

In general, one could consider either $\chi$ or $N_R$ dark matter in this model, though for reasons we now discuss the $N_R$ case may be more likely to lead to LFV signatures. Since the $\chi$ field has a non-zero hypercharge, its spin-independent direct-detection cross section is governed by tree-level $Z$-exchange. Assuming $\Omega_\chi$ saturates the relic density, the resulting direct detection cross section is too large to be phenomenologically viable~\cite{cerilli}. An exception occurs when  the mass-splitting between the real
and imaginary components of the neutral field, $S\equiv\mathrm{Re}\chi^0$ and $A\equiv\mathrm{Im}\chi^0$,  is greater than about 100 keV~\cite{idm1, idm2, idm3, idm4}. In this case, the dark matter particle can only scatter inelastically from the nuclei in the detector since the the $ZSS$ and $ZAA$ couplings vanish in contrast to the $ZSA$ interaction. For   $\delta =\ |m_S -m_{A}|\gsim 100$ keV, the inelastic scattering process mediated by tree-level $Z$-exchange is kinematically suppressed, yielding a phenomenologically viable direct detection cross section associated with the exchange of two gauge-bosons and/or the Standard Model Higgs. On the other hand, in order
for the $S$ or $A$ to saturate the relic density, its mass should be be order $10$TeV~\cite{smdm} or larger. These two considerations then imply that the coupling $|\tilde{\lambda}_{H\chi}|$ must be larger than  $\sim 10^{-4}$ since from  Eq.~(\ref{rimass}), one has  $\delta \approx \frac{6}{5}|\tilde{\lambda}_{H\chi}| \frac{v^2}{m_\chi}$.
 As we will see in Section IV, the product of  $\tilde{\lambda}_{H\chi} v^2$ and $Y^2$ sets the scale of neutrino masses, so that under these conditions $|Y|\lsim 10^{-4}$, implying that contributions from this model to lepton flavor violating processes will be too small to be observed.

For a considerably larger value of  $\delta$, a lower mass for $\chi_0$ can also be consistent with the dark matter relic density and direct detection limits. In this case, the scalar exchange and four-scalar interaction contributions to the annihilation process associated with the $\lambda^{1,2}_{H\chi}$ terms must be taken into account.  Cancellations can occur within the allowed parameter space between these contributions and those mediated by gauge bosons, leading to the correct DM relic density . Detailed studies have been carried out for the inert doublet model. It has been shown that
there exists two additional  mass ranges consistent with the relic density and direct detection limits: one with mass below 50 GeV~\cite{cerilli,50g} and another between $m_W$ and 160 GeV~\cite{160g}.
The same mechanism can also be applied to the model we study here. However, as with the inelastic dark matter scenario, an even larger $|\tilde \lambda_{H\chi}|$ is needed,  resulting in even smaller effects in LFV processes. Consequently, we defer a study of these lighter-mass $\chi^0$ scenarios to future work, where we will also consider the associated LHC phenomenology.


Consequently, we will take $N_R^0$  as the possible WIMP DM assuming that the relic density  is thermally produced in the standard $\Lambda \mathrm{CDM}$ model.
For a consistent model, at least two copies of $N_R$ are needed. Only the lightest $N_R^0$ is stable\cite{e-ma,other-neu} since the heavier  $N^0_{R j}$ can decay through the Yuakawa interaction by $N^0_{R j} \to L_{L_i} \chi$, followed by $\chi$ decays into a SM light lepton and the $N^0_R$. If $\chi$ mass is larger than that of the $N^0_{R j}$ as we assume here, there is a suppression factor due to the off-shellness of $\chi$. However, the next-to-lightest $N_R^{0j}$
is still unstable on cosmologically relevant timescales.

\begin{figure}
  \subfigure[]{\includegraphics[width=6cm]{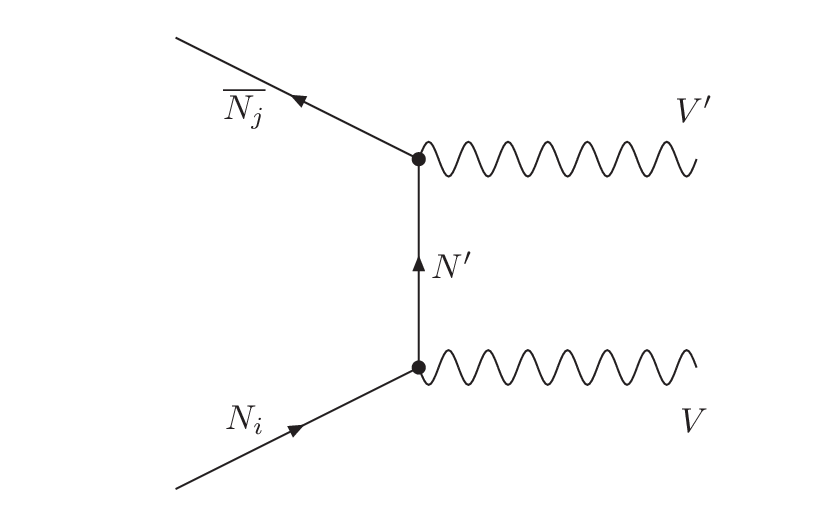}}\label{co-ann-s1}
  \subfigure[]{ \includegraphics[width=6cm]{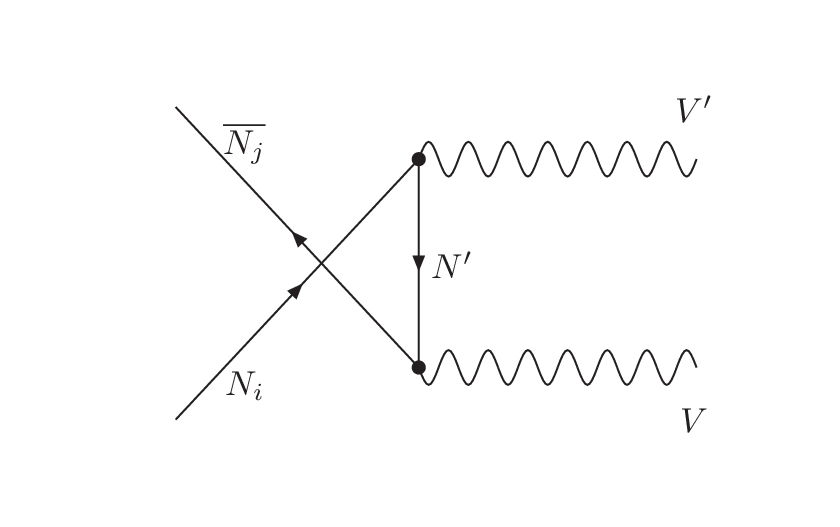}}\label{co-ann-s2}\\
  \subfigure[]{ \includegraphics[width=6cm]{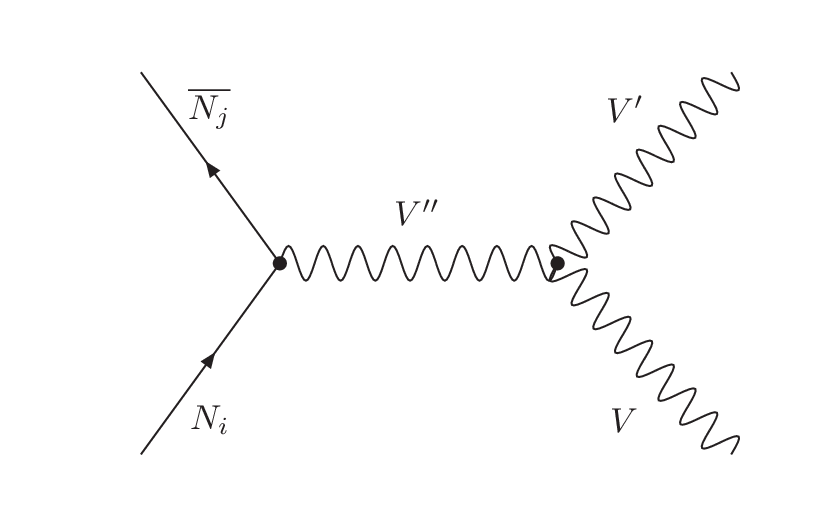}}\label{co-ann-s3}
    \subfigure[]{ \includegraphics[width=6cm]{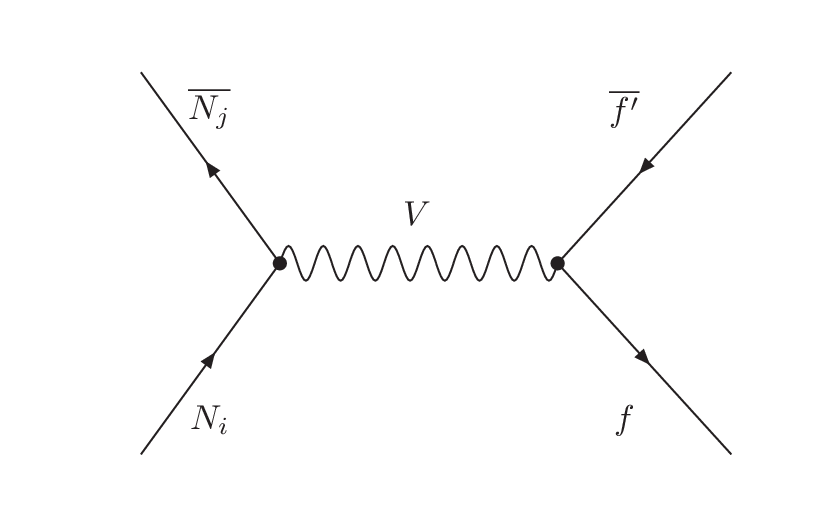}}\label{co-ann-s4}
  \caption{Feynman diagrams for co-annihilations. $V$ and $V'$, $f$ and $f'$ are appropriate SM gauge bosons and fermions. $N_{i,j}$ are component fields in $N_R$ with appropriate combinations of $i$ and $j$. For example the DM component is $N_i = N^0$ and its anti-particle $\overline N_j = N^{0c}$. A pair of DM annihilate themselves only through the t and u channels in the figure.}\label{co-ann}
\end{figure}

To produce the observed DM relic density $\Omega_{DM}h^2$, we require that the thermally averaged annihilation rate $\langle\sigma_A v\rangle$ to be $3\times 10^{-27}\mathrm{cm}^3s^{-1}/\Omega_{DM}h^2$.  In our model, $\sigma_A v$ is determined by co-annihilation\footnote{We use the term \lq\lq coannihilation" to denote processes in which the $N^0$ and either an $N^{0C}$ or one of the charged components of the multiplet annihilate into SM final states.} of a pair of $N$ into a pair of $W$ gauge bosons and other combinations as shown in Fig. \ref{co-ann}.  The interactions of $N_R$ with gauge bosons are contained in the kinetic term, $\bar N_R \gamma^\mu D_\mu N_R $.  Expanded in component forms, they are given by
\begin{eqnarray}
\nonumber
\Delta \mathcal{L}_{NNV}&=&(e A_\mu + g \cos\theta_W Z_\mu)(2 \bar N_R^{++}\gamma^\mu  N_R^{++} + \bar N_R^{+}\gamma^\mu  N_R^{+} - \bar N_R^{-}\gamma^\mu  N_R^{-} - 2 \bar N_R^{--}\gamma^\mu  N_R^{--})\\
\nonumber
&&+ g ( \sqrt{2} \bar N_R^{++}\gamma^\mu W^+_\mu N_R^{+} + \sqrt{3}\bar N_R^{+}\gamma^\mu W^+_\mu N_R^{0} \\
&& \qquad + \sqrt{3}\bar N_R^{0}\gamma^\mu W^+_\mu N_R^{-} + \sqrt{2}\bar N_R^{-}\gamma^\mu W^+_\mu N_R^{--}
+ h.c.),
\end{eqnarray}
which can be written, in terms of the four-component fields $N$, as
\begin{eqnarray}
\nonumber
\Delta \mathcal{L}_{NNV}&=& (e A_\mu + g \cos\theta_W Z_\mu)(2 \bar N^{++}\gamma^\mu  N^{++} + \bar N^{+}\gamma^\mu  N^{+})\\
&& + g (- \sqrt{2} \bar N^{++}\gamma^\mu W^+_\mu N^{+} - \sqrt{3}\bar N^{+}\gamma^\mu W^+_\mu N^{0}
+ h.c.)\; ,
\end{eqnarray}
where again we suppress the generational indices for simplicity.
Note that photon and $Z$ couplings to $N$ fields are vector-like.

Using the above interaction, to the leading order in $v^2$, one obtains~\cite{cerilli}
\begin{eqnarray}
\sigma_A v = {1\over 25\times 4 m^2_N}( c_s + c_p v^2)\;,\;\;\mbox{with}\; c_s = {1035\over 8\pi} g^4\;,\mbox{and}\; c_p = {1215\over 8 \pi} g^4\;.
\end{eqnarray}
$c_s$ and $c_p$ represent the strength of the S-wave and P-wave annihilation.

By fitting the relic density with the standard thermal relic density calculation, the DM mass can be determined.  Detailed calculations have been carried out in in Ref.\cite{cerilli}, so we will not repeat them here.  If one requires that $\Omega_{DM} h^2 = 0.110\pm 0.006$~\cite{pdg} to be saturated by thermal annihilation of $N_R$, the mass of the lightest $N_R$ component must have a mass of $4.4\pm 0.1$ TeV.  A more detailed analysis has been carried out in Ref.\cite{cerilli}. Taking into account the Sommerfeld effect that enhances the annihilation cross section, the DM mass is raised by about a factor of two with~\cite{cerilli} $m_N = (9.6\pm 0.2)$ TeV.  The model has some testable predictions for direct DM detection. DM interacts with quarks at one-loop level.  The direct DM detection cross section is predicted to be~\cite{cerilli} $10^{-44} \mathrm{cm}^2$ which is safely below the current upper limits from CDMSII and Xenon100 experiments, but can be tested at future~\cite{superCDMS} superCDMS and xenon-1ton experiments.

\begin{figure}
  \subfigure[]{\includegraphics[width=6cm]{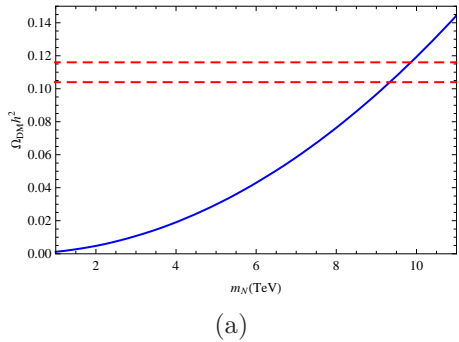}}
  \caption{$\Omega_{DM}h^2$ vs. $m_N$ with Sommerfeld effect taken into account.}\label{lhc}
\end{figure}

The DM mass of order 10 TeV makes it impossible to be directly probed at the LHC.  However, nature may choose to have several components of dark matter, with $N$ only produce a fraction of the total relic DM density~\footnote{In this case, the model needs to be further extended to accommodate the total DM relic density.}.  If so, the mass $m_N$ can be smaller.  To illustrate, we plot in Fig.\ref{lhc}, the relic density as a function of $m_N$. The relic density contribution from this model scales approximately as $m^2_N$.  Therefore, the relic density drops rapidly with decreasing $m_N$. We see that with $m_N\sim 1$ TeV, the relic density is already only about 1\% of the total. If the R$\nu$MDM scenario is going to play a significant role providing the DM relic density, it is unlikely that the LHC will be able to probe the heavy degrees of freedom our particular model realization.

\section{Neutrino masses}

\begin{figure}
\includegraphics[width=10cm]{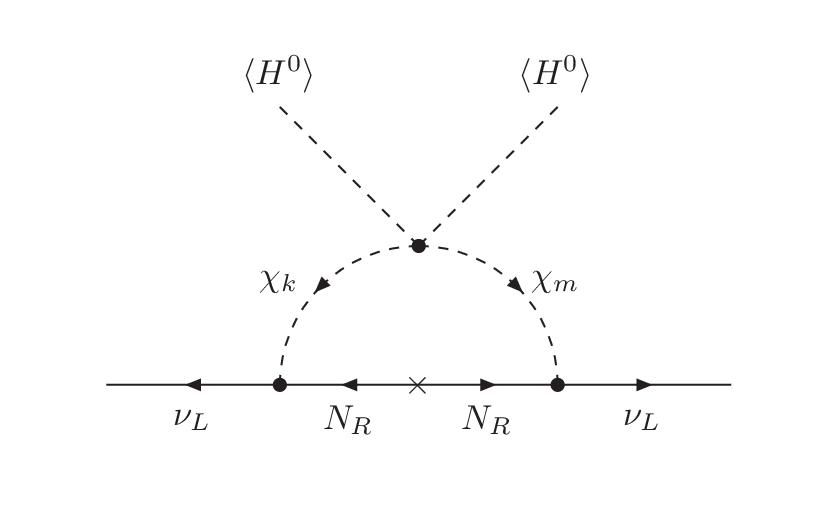}
  \caption{One loop Feynman diagram for neutrino mass generation. }\label{neu-mass}
\end{figure}

Light neutrinos do not have masses at  tree-level because the vev of $\chi$ is zero.  However, at one-loop level non-zero neutrino masses can be generated~\cite{e-ma} through the diagram shown in Fig.~\ref{neu-mass}. The relevant operators are the $N_R$ Yukawa interaction and the $\tilde \lambda_{H\chi}(H \chi)^2+\mathrm{h.c.}$ operator. None of the other quartic interactions involving $H$ and $\chi$ that appear in $V_{new}$ can lead to a lepton-number violating neutrino mass operator at one-loop since they contain a $\chi$ and $\chi^\dag$ pair.  The corresponding vertices in the diagram can be obtained from the following terms
\begin{eqnarray}
\bar L_L \chi N_R \to &&\bar \nu^i ({1\over \sqrt{5}} N_{Rj}^{++}\chi^{--} + \sqrt{{2\over 5}} N_{Rj}^+\chi^- + \sqrt{{3\over 5}} N_{Rj}^0\chi^0 - {2\over \sqrt{5}} N_{Rj}^- \chi^+ + N_{Rj}^{--}\chi^{++})\nonumber\\
&&+ \bar e^i (N_{Rj}^{++}\chi^{---} + {2\over \sqrt{5}} N_{Rj}^+\chi^{--} + \sqrt{{3\over 5}} N_{Rj}^0\chi^- - \sqrt{{2\over 5}} N_{Rj}^- \chi^0 + {1\over \sqrt{5}}N_{Rj}^{--}\chi^{+})\;,\nonumber\\
(H\chi)^2 \to && v^2({2\over \sqrt{5}}\chi^{++}\chi^{--} - {4\sqrt{2}\over 5} \chi^+\chi^- + {3\over 5} \chi^0\chi^0)\;,\nonumber\\
\end{eqnarray}
where the indices $i$ and $j$ are generation indices.

There are three pairs which can contribute to light neutrino masses in the loop. They are ($N^{++}$, $N^{--}$),  ($N^{+}$, $N^{-}$), and ($N^{0 c}$, $N^{0}$) with appropriate pairs of $\chi$ component fields in the loop.  The light neutrino mass matrix is given by
\begin{eqnarray}
&&(M_\nu)_{ll'} = {1\over {16} \pi^2} {14\over 5} {Y_{l i} Y_{l'i}\over m_{N_i}} \tilde \lambda_{H\chi} v^2 I(m_k,m_m, m_{N_i})\;,\nonumber\\
&&I(m_{\chi_k}, m_{\chi_m}, m_{N_i}) = {m^2_{N_i}\over m^2_{\chi_k} - m^2_{\chi_m}} \left ( {m^2_{\chi_k} \ln(m^2_{N_i}/m^2_{\chi_k})\over m^2_{N_i} - m^2_{\chi_k}}
- {m^2_{\chi_m} \ln(m^2_{N_i}/m^2_{\chi_m})\over m^2_{N_i} - m^2_{\chi_m}}\right )\;,
\end{eqnarray}
where $m_{\chi_{k,m}}$ and $m_{N_i}$ are the masses of the scalars, and fermions in the loop.

Neglecting the mass splitting of the component fields in $\chi$ (with a common mass $m_\chi$), $I(m_k,m_m,m_N)$ becomes
\begin{eqnarray}
\tilde I(z) = \frac{(1-z)z+z^2\ln z}{(1-z)^2}\;,
\end{eqnarray}
where $z = m^2_N/m^2_\chi$. In the limit $m_N = m_\chi$, $\tilde I(1) = 1/2$.

Diagonalizing $M_\nu$, one obtains the Pontecorvo$-$Maki$-$Nakagawa$-$Sakata (PMNS) mixing matrix $V_\mathrm{PMNS}$ for the charged current interaction involving leptons, $M_\nu = V_{\text{PMNS}} \hat M_\nu V_{\text{PMNS}}^T$.  Here $\hat M_\nu$ is the diagonalized light neutrino mass matrix. The signs of the eigen-masses can be normalized to be positive  by appropriate chiral rotations of the heavy neutrino fields. A general solution for $Y$ can be written as
\begin{eqnarray}
Y=\bigg(\frac{v^2}{{16} \pi^2}\frac{14}{5}\tilde{\lambda}_{H\chi}\bigg)^{-1/2} V_{\text{PMNS}} \hat{M}_\nu^{1/2}O\hat{M}_N^{1/2} \hat{\tilde I}^{-1/2},
\label{yukawa}
\end{eqnarray}
where $\hat M_n$ and $\hat {\tilde I}$ are diagonal matrices with $\hat M_N = diag(m_{N_1}, m_{N_2},...)$ and  $\hat {\tilde I} = diag(\tilde I(z_1), \tilde I(z_2), ...)$. $O$ satisfies $OO^T=I$ and can be complex in general.

From the above one finds that in order to have at least two massive light neutrinos, one requires at least two copies of $N_R$ fields. With two $N_R$, one of the light neutrino masses is zero.  If it turns out that all the three light neutrinos are massive, at least three $N_R$ fields must be introduced.
In the presence of multiple, non-degenerate $N_R$ generations, the lightest one  plays the dominant role in setting the mass scale for neutrinos.  If they are almost degenerate, all will contribute significantly. In all these cases, the DM mass plays the role in setting the lower scale of the heavy particles in the RSSM.

With three degenerate $m_{N_{1,2,3}} = m_N$, one can easily find solutions of $Y$ which give the correct neutrino mixing pattern and also their masses in the allowed ranges. As an illustration, we take $O=I$ and $\tilde{\lambda}_{H\chi}=10^{-7}$, $m_N=m_\chi =9.6\;\text{TeV}$, $V_\mathrm{PMNS}$ to be the tri-bimaximal form~\cite{tri-bimaximal}, and obtain sample Yukawa couplings $Y_\text{NH,IH}$ with the light neutrinos in a normal mass hierarchy $(0,0.009,\;0.05)$ eV and an inverted hierarchy $(0.05, 0.051,\;0.01)$ eV. We have the corresponding Yukawa couplings given by
\begin{eqnarray}
Y_{\text{NH}}=\left(
\begin{array}{ccc}
0&0.032&0\\
0&0.032&0.095\\
0&0.032&-0.095\\
\end{array}\right)\;,\;
Y_{\text{IH}}=\left(
\begin{array}{ccc}
-0.11&0.078&0\\
0.055&0.078&0.046\\
0.055&0.078&-0.046\\
\end{array}\right)\;.
\end{eqnarray}

For different choices of $O$, the resulting $Y$ are also different, reflecting the fact that the number of parameters is larger than the number of constraints. The key point, however,  is that for a generic choice of $O$ one can have large Yukawa couplings that may in turn have some implications for other leptonic flavor violating phenomena. Importantly, the Yukawa coupling $Y$ scales as $\tilde \lambda_{H\chi}^{-1/2}$. By adjusting the size of $\tilde \lambda_{H\chi}$, one can change the overall magnitude of the Yukawa couplings. In the above a small number for $\tilde \lambda_{H\chi}$ has been used. If one sets this coupling to be zero, the model has a global $U(1)$ lepton number symmetry. In that sense a small number for $\tilde \lambda_{H\chi}$ is technically natural. Indeed, explicit inspection of the one-loop corrections shows that this operator does not mix with the others in $V_{new}$ under renormalization. Moreover, as emphasized earlier, the  a non-negative $\tilde \lambda_{H\chi}$ will remain so after RG running.  So the scenario with small $\tilde \lambda_{H\chi}$ is consistent with vacuum stability considerations.

As discussed above, for $|\tilde \lambda_{H\chi}|$ having a significantly larger magnitude (of order $\sim 10^{-4}$), one may also be able to consider inelastic $\chi$ dark matter, since in this case the associated mass-splitting between the real and imaginary parts of $\chi^0$ will be sufficiently large to suppress the inelastic direct detection cross section. However, from Eq.~(\ref{yukawa}) and the overall scale of neutrino masses, we find that the magnitude of the Yukawa couplings must be generically $\lsim 10^{-4}$. As we will see in Section \ref{sec:lfv}, the corresponding effects on lepton flavor violating observables would then be too small to be observable in the next generation of LFV experiments. Since our emphasis here falls on the possible signatures of the R$\nu$MDM scenario and LFV, we will not elaborate further on the large inelastic $\chi$ dark matter possibility.

\section{LFV: present constraints and future probes}
\label{sec:lfv}

With the dark matter mass determined  $~10$ TeV, this model has zero chance to be tested by directly producing the new heavy degrees of freedom at the LHC. As  discussed earlier,  a $N_R$ mass lighter than one TeV  that could in principle lead to LHC discovery would imply undersaturation of the relic density by a factor of 100 -- too small to play a significant role for dark matter.  As an alternative probe, we consider charged  lepton flavor violating processes assuming that this model saturates the DM relic density. Indeed, the light neutrino masses arise at one-loop order and have a magnitude governed by $M_N$ and $M_\chi$ and by products of the Yukawa couplings $Y$ and ${\tilde\lambda}_{H\chi}$, whereas loop-induced LFV amplitudes are independent of  ${\tilde\lambda}_{H\chi}$, Consequently, it is possible that that Yukawa couplings can be sufficiently large to generate observable LFV signatures while remaining consistent with the scale of light neutrino masses. In what follows, we show that the model can lead to a large $\mu \to e \gamma$ branching ratio, comparable in magnitude to the recent limit reported by the MEG collaboration\cite{MEG} and its expected  future sensitivity\cite{meg} as well as a large $\mu -e$ conversion rate that may be accessible with the Mu2E \cite{Ref:Mu2E}, COMET\cite{Ref:COMET}, and PRISM\cite{Ref:PRISM} experiments.  The LFV decays $\tau \to \mu (e) \gamma$ can also provide additional information.

\noindent{\bf Constraint from $\mu \to e \gamma$}

The decay amplitudes for $\ell_i\to \ell_j \gamma$ can in general be parameterized as
\begin{eqnarray}
\label{eq:dipole}
\mathcal{M}=\epsilon^{*\mu}\bar{\ell}_j i\sigma_{\mu\nu}q^\nu(\tilde A_R P_R+ \tilde A_L P_L) \ell_i\;.\label{EMamp}
\end{eqnarray}
Neglecting small mass splitting of component fields within a given $N_R$ multiplet, we obtain
\begin{eqnarray}
&&\tilde A_R = \frac{ m_i e }{32\pi^2m_{\chi}^2}\ \sum_k Y_{ik}^*Y_{jk}\, F(z_k)\;,\nonumber\\
&&F(z) = P_\chi \frac{2z^3+3z^2-6z+1-6z^2\ln z}{6(1-z)^4} - P_N\frac{z^3-6z^2+3z+2+6z\ln z}{6(1-z)^4}\;,\nonumber\\
\end{eqnarray}
where $z_k = m^2_{N_k}/m^2_{\chi}$. Summarizing all one loop contributions, we have $P_\chi=-5$ and $P_N=2$.
$\tilde{A}_L$ can be obtained by replacing $m_i$ by $m_j$ in the above.

Note that the $\mu \to e \gamma$ amplitude does not depend on the parameter $\tilde \lambda_{H\chi}$. As emphasized earlier, this feature allows the possibility of having small $\tilde \lambda_{H\chi}$, but large $Y$ to satisfy constraints from neutrino masses and mixing, and to have a large $\mu \to e \gamma$ branching ratio. Moreover, since the DM relic density is governed by the gauge rather than Yukawa interactions, one still requires  knowledge of the  $Y_{ij}$ to determine $l_i \to l_j \gamma$, even with the scale of heavy degrees of freedom fixed by the DM relic density.  It is clear that the known constraints on light neutrino masses and mixing matrix provide important information, yet they do not complete fix the products  $Y_{ik}^*Y_{jk}$. Consequently, one may consider the LFV searches as independent probes of the Yukawa structure of the model, assuming sufficient sensitivity. To illustrate and to simplify the analysis, we will take the heavy degrees of freedom to be degenerate, i.e., $m_N=m_{N_i}$. Note that the mass splittings between generations so as to allow for a one-species DM scenario can be sufficiently small that we may neglect them here. Similarly, the splittings within a generation induced by weak radiative corrections are negligible for this purpose.  We will then work with the partial branching ratio defined as
$ \overline{\text{BR}}(l_i\to l_j\gamma)\equiv \text{Br}(l_i \to l_j \gamma)/\text{Br}(l_i \to l_j \nu\bar \nu)$.

\begin{figure}
  \includegraphics[width=8cm]{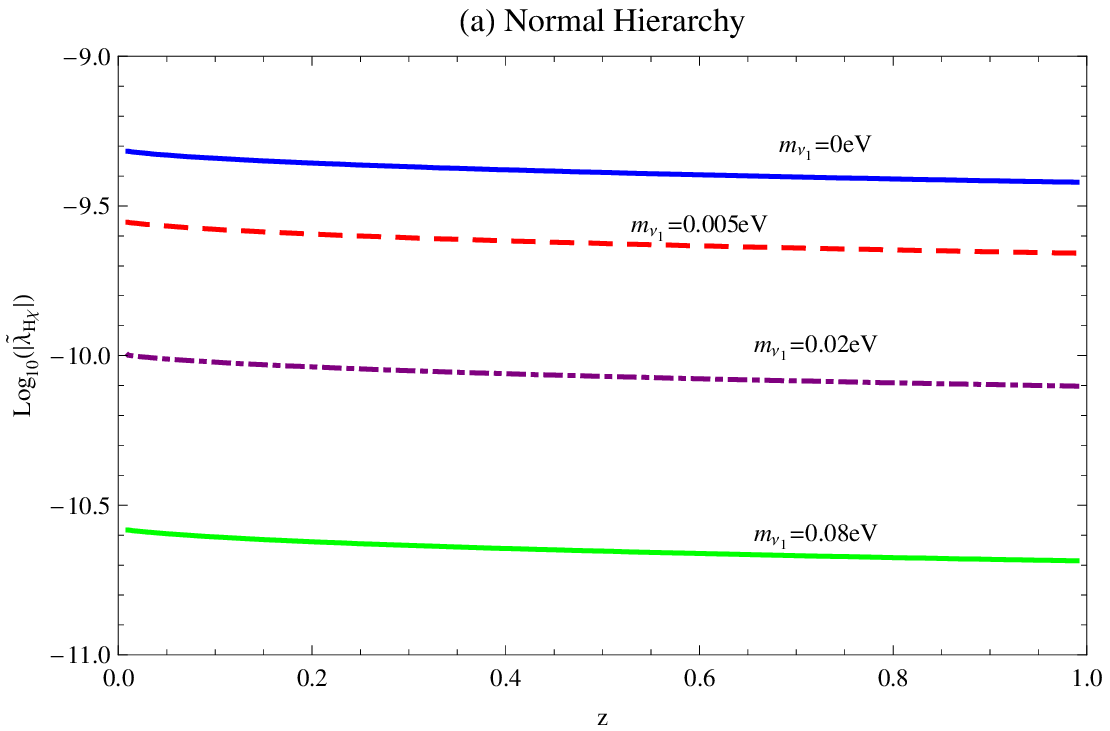}
  \includegraphics[width=8cm]{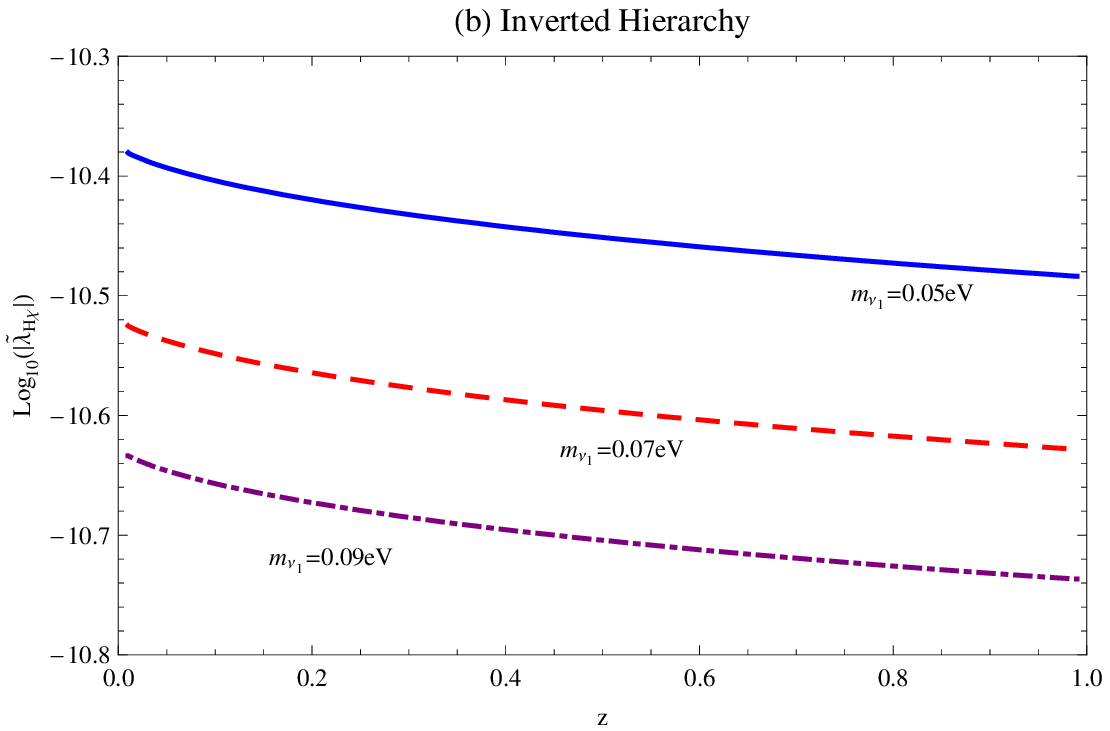}\\
  \caption{$\log_{10}|\tilde{\lambda}_{H\chi}|$ vs. $z=m^2_N/m^2_\chi$ for (a) normal hierarchy, and (b) inverted hierarchy for different choices of $m_{\nu_1}$. The  curves are obtained using the recent MEG upper limit $\text{Br}(\mu\to e\gamma)=2.4 \times10^{-12}$. }\label{Figmu2egamma}
\end{figure}

Although there presently exists no evidence for any decays $l_i \to l_j \gamma$, impressive bounds have been obtained in many cases. For many years, the most stringent on $\overline{\text{BR}}(\mu \to e\gamma)$ is $1.2\times 10^{-11}$\cite{pdg}. Very recently, the MEG collaboration has obtained a better upper limit with\cite{MEG} $\overline{\text{BR}}(\mu \to e\gamma) < 2.4 \times 10^{-11}$ at  90\% C.L. One way to ascertain whether our model can have testable
consequences is to see if with known constraints, the model can produce partial branching ratios close to the current experimental bounds.
To this end, we first vary $m_{\nu_1}$ from $0\;\text{eV}$ to $0.08\;\text{eV}$ for the normal hierarchy, and $0.05\,\text{eV}$ to $0.09\,\text{eV}$ for the inverted hierarchy, with the light neutrino mass square differences, $\Delta m_{21}^2=(7.59\pm0.21)\times10^{-5}\mathrm{eV}^2$~\cite{Aharmim:2009gd}, and $|\Delta m_{32}^2|=(2.43\pm0.13)\times10^{-3}\mathrm{eV}^2$~\cite{Adamson:2008zt} fixed from experiments. These choices also satisfy the cosmological constraint $\sum_i m_{\nu_i}<0.28\;\text{eV}$\cite{cos-mass}. We then determine the allowed parameter space assuming that this model saturates the $\overline{\text{Br}}(\mu\to e\gamma)$ experimental upper bound $2.4\times 10^{-12}$. We then obtain the constraint between $|\tilde{\lambda}_{H\chi}|$ and $m_\chi$ with Eq.~(\ref{yukawa}) applied as well. The results are shown in Fig.~\ref{Figmu2egamma}.

We comment that in order to have a large branching ratio for $\mu \to e \gamma$ close to the upper bound in both normal and inverted hierarchies with $m_{\nu_1}$ larger than 0.08 eV, the parameter $\tilde \lambda_{H\chi}$ is smaller than $10^{-11}$. In this regime, the elements of the Yukawa matrix $Y$ are of order a few, which may be close to the non-perturbative region wherein the results may become less accurate compared with smaller $m_{\nu_1}$.  But for $m_{\nu_1}$ less than 0.08 eV, the  perturbative predictions should be reasonably reliable.

\begin{figure}
  \includegraphics[width=8cm]{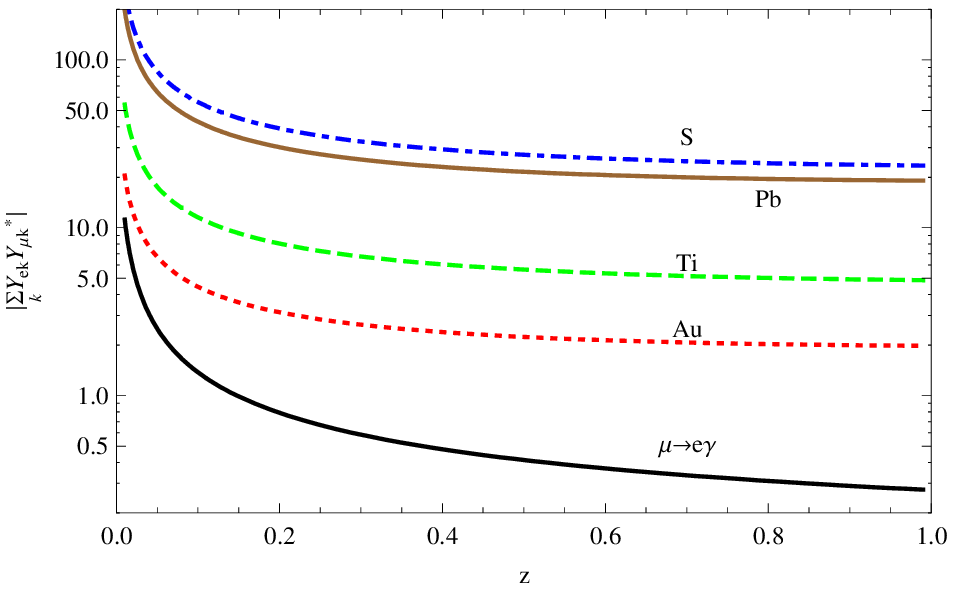}
  \includegraphics[width=8cm]{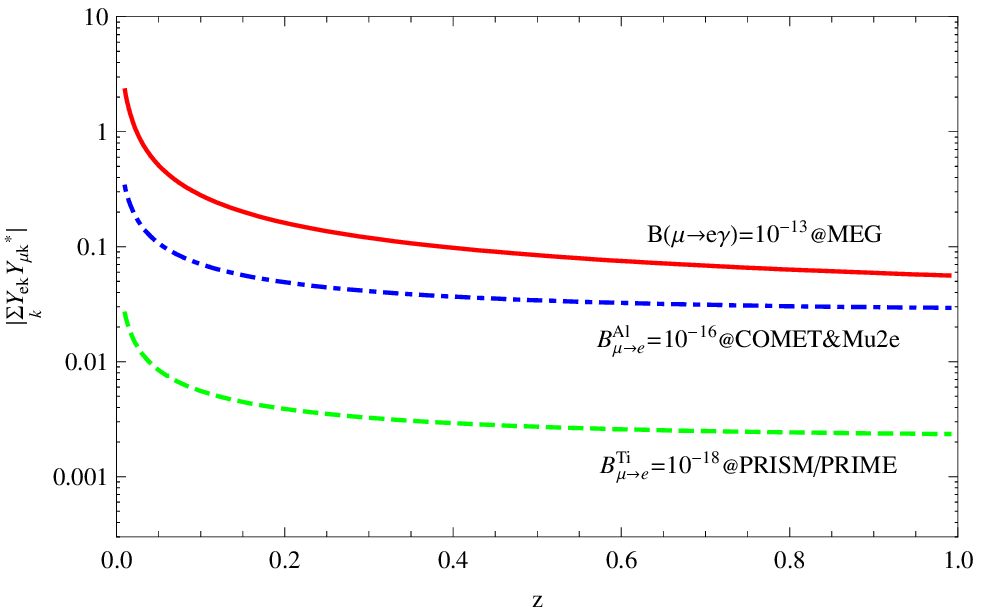}\\
  \caption{Comparison of present (left panel) and prospective (right panel) constraints from $\mu -e$ conversion on various nuclei and $\mu \to e \gamma$.}\label{mu-e-conversion}
\end{figure}

In Fig.\ref{mu-e-conversion}, we also show the constraint on the relevant product of
Yukawa couplings using the current experimental bounds on $\mu \to e \gamma$ and $ \mu - e $ conversion. Present constraints from $\mu \to e \gamma$ as well as searches for $\mu\to e$ conversion in nuclei (see below) are given in the left panel, while the sensitivities of prospective experiments are shown in the right panel. We see that for $m_\chi$ not too much larger than $m_N$, that is $z = (m_N/m_\chi)^2$ close to 1 the combination
of the Yukawa coupling $\sum_k Y_{e k}Y^*_{\mu k}$ is presently constrained to be less than 1 (left panel). The upper curve in the right panel gives the expected sensitivity of the MEG experiment. We observe that even with the assumption that the Yukawa couplings are less than order one, the model may have observable effects in $\mu \to e \gamma$.


\noindent{\bf Constraints from $\mu - e$ conversion}

In the case of  $\mu - e $ conversion,  new operators contribute to the LFV amplitude; in addition to the dipole interaction (\ref{eq:dipole}) one also must include the charge radius operator. In general there is also a $Z$-penguin contribution. However, due to the vector-like coupling of $Z$ to $N$, this operator vanishes for the same reason that $\mu e\gamma$ charge operator vanishes, a feature that we have verified explicitly. This situation differs, for example,  from that in the SM with four generations, where $Z$-penguin dominates the contribution~\cite{four}. Moreover, the $Z$ dipole and charge radius-induced  $\mu - e$ conversion amplitude is suppressed by a factor of $m^2_\mu/m^2_Z$. Consequently, all $Z$-exchange contributions can be safely neglected.

The photon charge radius-generated
quark level $\mu -e $ conversion amplitude given by
\begin{eqnarray}
L &=& {G_F\over \sqrt{2}} {s^2_W \over 2 \pi^2} {m^2_W\over m^2_\chi}\, \sum_k Y_{ek}Y^*_{\mu k} \bar e \gamma_\mu P_L \mu [P_\chi G_\chi(z_k) + P_N G_N(z_k)] \sum_q Q_q \bar q \gamma^\mu q\;,\label{mu-ec}
\end{eqnarray}
where $Q_q$ is the electric charge of quark q in unit $e$, $s_W=\sin\theta_W$ and $c_W = \cos\theta_W$, and
\begin{eqnarray}
&&G_\chi(z) = {2-9z+18z^2-11z^3+6z^3\ln z\over 36(1-z)^4}\;,\nonumber\\
&&G_N(z) = {-16+45z-36z^2+7z^3+6(3z-2)\ln z\over 36(1-z)^4}\;.
\end{eqnarray}

Several groups have performed searches for $\mu - e$ conversion~\cite{pdg}.  In Table  \ref{Tab:mu-e} we list results  for the conversion-to-capture ratio
\begin{eqnarray}
B^A_{\mu \to e} = {\Gamma^A_{conv}\over \Gamma^A_{capt}} =
{\Gamma(\mu^- + A(N,Z) \to e^- +A(N,Z))\over \Gamma(\mu^- + A(N,Z) \to \nu_\mu + A(N+1, Z-1)}\;,
\end{eqnarray}
where $A$ denotes the atomic number .

To obtain $\mu \to e$ conversion rates on different nuclei from the FCNC interaction in the above, we start with the effective four-fermion $e\mu q{\bar q}$ operators that may contribute. Following the notation of Ref.~\cite{kitano}  we have
\begin{eqnarray}
L_\mathrm{eff} &=& -{4 G_F\over \sqrt{2}} \left [m_\mu \bar e \sigma^{\mu\nu} (A_R P_R + A_L P_L) \mu F_{\mu\nu}  + h.c.\right ]\nonumber\\
& - & {G_F\over \sqrt{2}} \left [ \bar e (g_{LS(q)}  P_R  + g_{RS(q)} P_L )\mu\, \bar q q  + \bar e (g_{LP(q)}  P_R  + g_{RP(q)} P_L )\mu\, \bar q \gamma_5 q+ h.c.\right ]\nonumber\\
& - & {G_F\over \sqrt{2}} \left [ \bar e (g_{LV(q)} \gamma^\mu P_L  + g_{RV(q)} \gamma^\mu P_R )\mu\, \bar q\gamma_\mu  q  + \bar e (g_{LA(q)}  \gamma^\mu P_L  + g_{RA(q)} \gamma^\mu P_R )\mu\, \bar q \gamma_\mu \gamma_5 q + h.c. \right ]\nonumber\\
&-& {G_F\over \sqrt{2}} \left [ {1\over 2}\bar e (g_{LT(q)} \sigma^{\mu\nu} P_R  + g_{RT(q)} \sigma^{\mu\nu} P_L )\mu\, \bar q\sigma_{\mu\nu}  q + h.c.\right ]\;.
\end{eqnarray}

Comparing with Eq.~\ref{mu-ec}, we have, at the one-loop level, $g_{(L,R)(S,P,A,T)(q)} = 0$, $g_{RV(q)}=0$,  and
\begin{eqnarray}
&&A_R = \frac{\sqrt{2}}{8} {\tilde A_R \over G_F m_\mu}\;,\;\; A_L = {m_e \over m_\mu} A_R\;,\\
&&g_{LV(q)} = -{s^2_W\over 2 \pi^2}\, \sum_k{m^2_W\over m^2_{N_k}} Q_q Y_{ek}Y^*_{\mu k} z_k\left[P_\chi G_\chi(z_k) + P_N G_N(z_k)\right]\;.
\label{arl}\nonumber\\
\end{eqnarray}
The contribution from $A_L$ can be neglected compared with that from $A_R$.

Theoretical efforts have been made by several group to calculated these matrix elements. We will use the results in Ref.\cite{kitano} for a consistent calculation.
The resulting expression for quantity $B^A_{\mu \to e}$ measuring the leptonic FCNC effect in $\mu\to e$ conversion is proportional to
\begin{eqnarray}
|A_R D + \tilde g^{(p)}_{LV} V^{(p)} + \tilde g^{(n)}_{LV} V^{(n)}|^2\;,
\end{eqnarray}
where $D(A)$, $V^{(p)}(A)$ and $V^{(n)}(A)$ are overlap integrals as a function of atomic number~\cite{kitano} and
\begin{eqnarray}
\tilde g^{(p)}_{LV} = 2 g_{LV(u)}+g_{LV(d)}\;,\;\;\tilde g^{(n)}_{LV} = g_{LV(u)}+ 2 g_{LV(d)}\;.
\end{eqnarray}
Combining these, we obtain
\begin{eqnarray}
{B^A_{\mu\to e}\over B(\mu \to e\gamma)} = R^0_{\mu\to e}(A) \left | 1 + {\tilde g^{(p)}_{LV} V^{(p)}(A)\over A_R D(A)} + {\tilde g^{(n)}_{LV} V^{(n)}(A)\over A_R D(A)}\right |^2\;,
\end{eqnarray}
where
\begin{eqnarray}
R^0_{\mu\to e}(A) = {G^2_F m^5_\mu \over 192 \pi^2 \Gamma^A_{capt}}|D(A)|^2\;.
\end{eqnarray}
The relevant parameters are listed  in Table \ref{Tab:mu-e}, which are
evaluated using method I in Ref.\cite{kitano}.
\begin{table}[tb]
\begin{tabular}{|l||l|l|l|l|l|}
\hline
A &upper limit& $D$(A) & $V^{(p)}$(A) & $V^{(n))}$(A) & $R^0_{\mu\to e}$(A) \\
\hline
${}^{27}_{13}$Al & -- & 0.0362 & 0.0161 & 0.0173 & 0.0026 \\
${}^{32}_{16}$S  & $7.0\times 10^{-11}$& 0.0524 & 0.0236 & 0.0236 & 0.0028 \\
${}^{48}_{22}$Ti & $4.3\times 10^{-12}$& 0.0864 & 0.0396 & 0.0468 & 0.0041 \\
${}^{197}_{79}$Au &$7.0\times 10^{-13}$ & 0.189  & 0.0974 & 0.146  & 0.0039 \\
${}^{208}_{82}$Pb & $4.6\times 10^{-11}$& 0.161  & 0.0834 & 0.128  & 0.0027 \\
\hline
\end{tabular}
\caption{The relevant parameters for $\mu - e$ conversion processes, which are evaluated
by using method I in Ref.\cite{kitano}. The upper limits are their 90\% c.l. values.}
\label{Tab:mu-e}
\end{table}

The constraints on $\sum_k Y_{ek}Y^*_{\mu k}$ from $\mu -e $ conversion on various nuclei and $\mu \to e\gamma$ are shown in Fig.\ref{mu-e-conversion} on the left panel assuming that $N_k$ are degenerate. We see that among the current available experimental limits, $\mu \to e \gamma$ gives the strongest constraints. The current bounds on $\mu -e $ conversion do not provide significant constrains on the parameter since the Yukawa couplings can still be much larger than of order 1. On the other hand, the sensitivities of prospective future $\mu-e$ conversion searches Mu2E\cite{Ref:Mu2E}/COMET\cite{Ref:COMET} and PRISM\cite{Ref:PRISM} exceed that of the MEG experiment.


\noindent{ \bf Effects on $\tau \to \mu (e) \gamma$}

\begin{figure}
  \includegraphics[width=8cm]{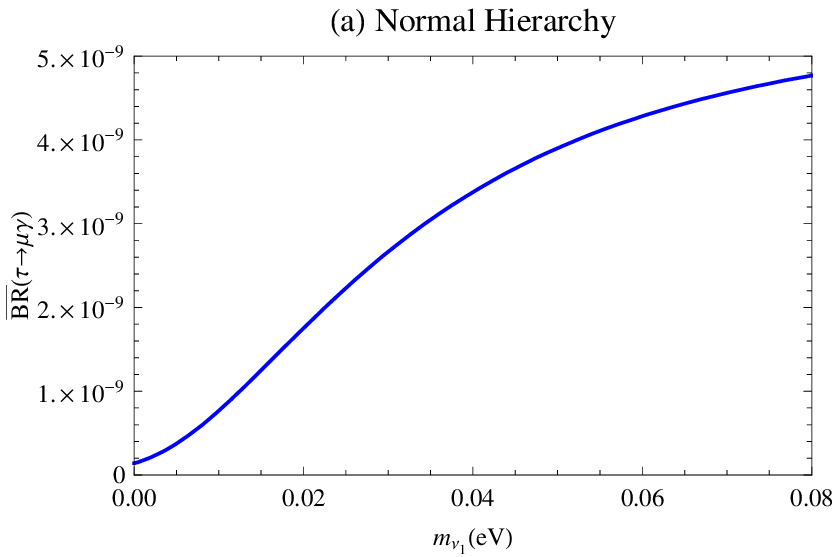}
  \includegraphics[width=8cm]{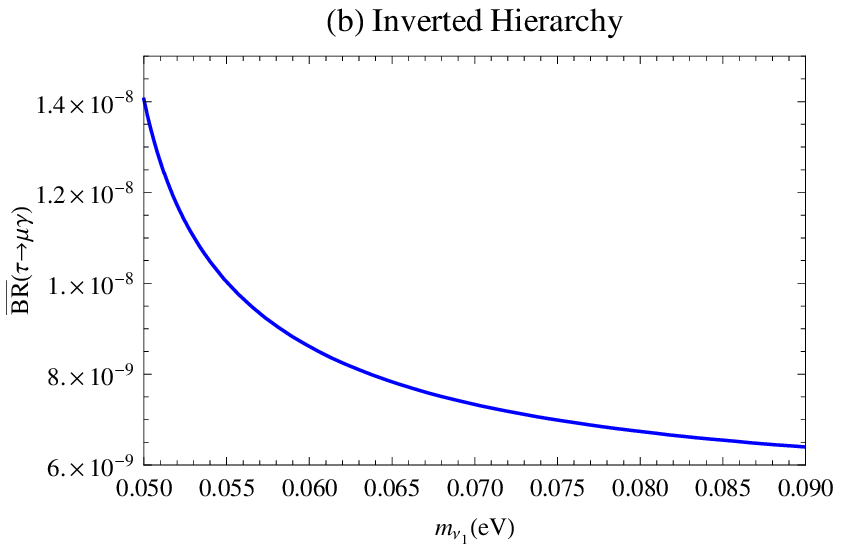}\\
  \caption{$\overline{\text{Br}}(\tau\to \mu\gamma)$ as a function of $m_{\nu_1}$ in (a) normal hierarchy, and (b) inverted hierarchy.}\label{Figtau2mugamma}
\end{figure}

\begin{figure}
  \includegraphics[width=8cm]{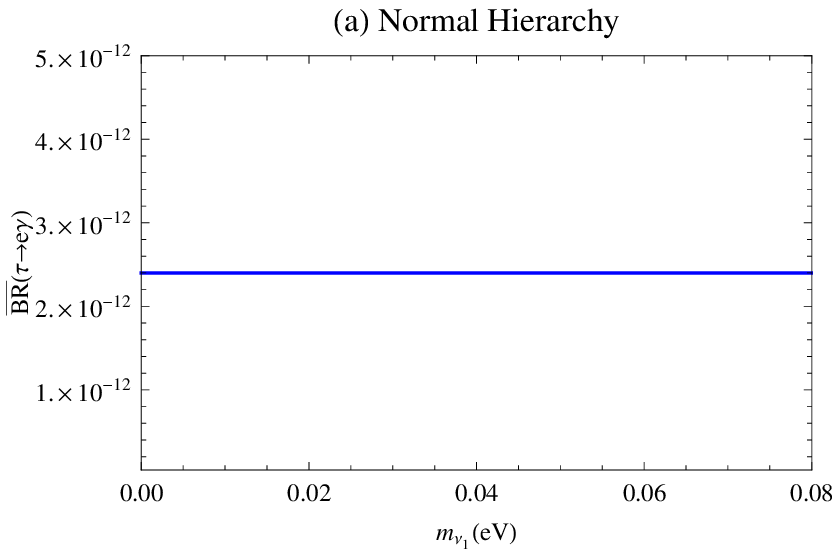}
  \includegraphics[width=8cm]{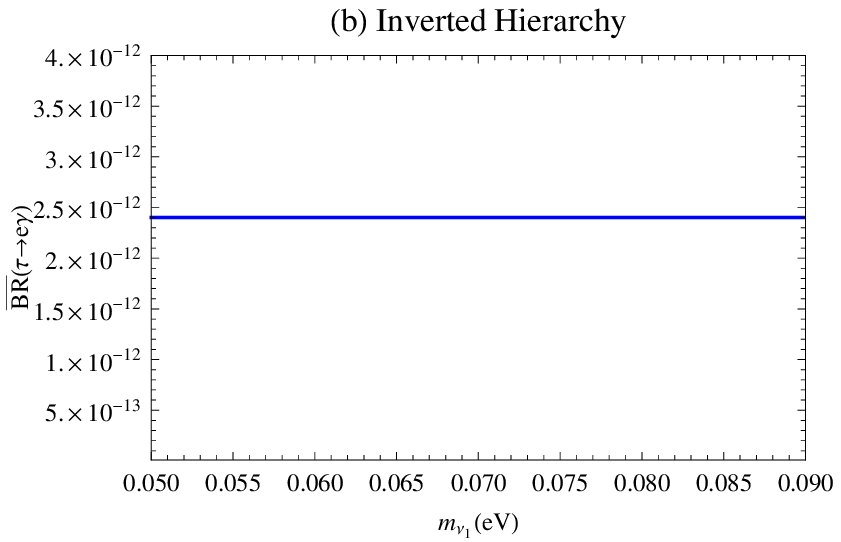}\\
  \caption{$\overline{\text{Br}}(\tau\to e \gamma)$ as a function of $m_{\nu_1}$ in (a) normal hierarchy, and (b) inverted hierarchy.}\label{Figtau2egamma}
\end{figure}

We now study the allowed ranges for $\tau \to \mu (e) \gamma$. The current experimental bounds are: $\overline{\text{BR}}(\tau \to \mu \gamma) < 2.5\times10^{-7}$ and $\overline{\text{BR}}(\tau \to e\gamma) < 1.8\times10^{-7}$~\cite{pdg}. For illustration we will consider the case for degenerate $N_i$ with $ m_N = 9.6$ TeV.
In this case once the value of $\overline{\text{Br}}(\mu\to e\gamma)$ is fixed, the entries in the Yukawa matrix $Y$ are also fixed in terms of $m_{\nu_1}$. Therefore, predictions for $\overline{\text{Br}}(\tau\to \mu\gamma)$ and $\overline{\text{Br}}(\tau\to e\gamma)$ can be made as a function of $m_{\nu_1}$. We show the results in Fig.~\ref{Figtau2mugamma} and Fig.~\ref{Figtau2egamma}. From Fig.~\ref{Figtau2mugamma}, we see that $\overline{\text{Br}}(\tau\to \mu\gamma)$ increases, and decreases when $m_{\nu_1}$ increases for the normal and inverted hierarchy cases, respectively. The predicted partial branching ratio can be about $10^{-8}$ which may be probed by future experiment at the super B factory\cite{superB}. For $\tau \to e \gamma$ the partial branching ratio is predicted to be much smaller than the current experimental limit making it difficult to test the model using this process.

In the examples given above, we have used the tri-bimaximal form for the mixing matrix $V_{PMNS}$. We have also checked cases with $\theta_{13}$ set to be non-zero. We find a similar pattern, but slightly smaller partial branching ratios. Because in our model, we have approximately,
$\overline{\text{BR}}(l_i \to l_j\gamma)= \overline{\text{BR}}(\mu \to e\gamma)
|[(V_{PMNS})^\dagger \hat{M}_\nu(V_{PMNS})^*]_{ji}|^2/|[(V_{PMNS})^\dagger \hat{M}_\nu(V_{PMNS})^*]_{12}|^2$, with a non-zero $\theta_{13}$ the denominator of the right-hand side becomes larger. This leads to smaller branching ratios for $\tau \to \mu(e)\gamma$ compared with the cases discussed earlier.

\noindent{ \bf Flavor diagonal lepton magnetic moments}

Before ending this section, we would like to make some comments on magnetic dipole moments of $\mu$ and neutrinos.
The anomalous magnetic dipole moment of muon $a_\mu$ can be generated at one loop level in our model. The current value of difference between experiment and SM is~\cite{pdg} $\Delta a_\mu=(255\pm63\pm49)\times10^{-11}$. The new contribution for $a_{l = e, \mu, \tau}$ in this model is given by $\Delta a_\ell= (m_\ell^2/16\pi^2m_N^2)zF(z)\sum_k |Y_{\ell k}|^2$.
Again, for a given $m_{\nu_1}$ with fixed $\mu \to e\gamma$ branching ratio, $\Delta a_\ell$ can be obtained. The result is much smaller than the current constraint given previously. For example, $\Delta a_\mu=-6.8\times10^{-11}$ when $m_{\nu_1}=0.08\;\text{eV}$ with normal hierarchy, and $\Delta a_\mu=-1.5\times10^{-11}$ when $m_{\nu_1}=0.05\;\text{eV}$ with inverted hierarchy.

Neutrinos can also have magnetic dipole moments. For Majorana neutrinos only transition magnetic moments $\mu_{ij}$ may be non-vanishing.
In our model, the one loop level calculation, obtained by attaching an external photon to the charged lines in Fig.\ref{neu-mass} in all possible ways, results in
a vanishing $\mu_{ij}$. It can only be generated at higher orders. We have estimated the two loop contribution and found that $|\mu_{ij}|$ is of order $10^{-20}\mu_B$.  This is much smaller than current experimental upper bound is $\mu_\nu<0.32\times10^{-10}\mu_B$ at $\text{CL}=90\%\,$~\cite{Beda:2009kx}.

\section{Summary}
In this work, we have constructed a scenario -- the R$\nu$MDM model --  relating the \lq\lq minimal" dark matter and radiative seesaw neutrino mass scales. An important feature  is that this model does not impose any beyond-the-SM gauge symmetry.  It contains, in additional to the SM particles, a Majorana fermion multiplet $N_R$ and a scalar multiplet $\chi$ which transform respectively as $(1,5,0)$ and  $(1,6,-1/2)$ under the SM gauge group $SU(3)_C\times SU(2)_L\times U(1)_Y$.  The lightest new component of the  $N_R$ is electrically neutral and does not decay into SM particles, making it a natural candidate for
DM. The DM can (co-)annihilate into SM particles through electroweak interactions.  To produce the correct DM thermal relic density, the DM mass is determined to be in the range 9 to 10 TeV.  This scale also sets the lower limit for the scalar $\chi$ mass scale which, in combination with $N_R$, generates light neutrino masses through radiative seesaw mechanism.  The large DM mass of order 10 TeV makes it impossible to directly detect the new particles at the LHC. This model, however, predicts that there is a sizable direct DM detection cross section of order $10^{-44}$ cm$^2$ which can be tested in future experiments, such as superCDMS, and Xenon-1ton. This scenario can also accommodate present neutrino data while leading to a large $\mu\to e\gamma$ branching ratio that can be probed by the MEG experiment.  Present constraints obtained from $\mu - e$ conversion searches are weaker than those from $\mu \to e\gamma$.  However, future $\mu -e$ conversion experiments on $Al$ and $Ti$ can provide much more stringent constraints.  This model also can imply a sizeable $\tau \to \mu \gamma$ decay rate that may be observable at super B factories.

\acknowledgments \vspace*{-1ex}
Y.C, X.G.H and L.H.T were supported in part by NSC, NCTS,  NNSF, SJTU Innovation Fund for Postgraduates and Postdocs. M.J.R-M was supported in part by Department of Energy contract DE-FG02-08ER41531 and by the Wisconsin Alumni Research Foundation.


\bigskip




\begin{thebibliography}{0}

\bibitem{pdg} K. Nakamura (Particle Data Group), J. Phys. {\bf G37}, 075021(2010).

\bibitem{silk} G.~Bertone, D.~Hooper and J.~Silk,
  Phys.\ Rept.\  {\bf 405}, 279 (2005)
  [arXiv:hep-ph/0404175].



\bibitem{singlet}
  V.~Silveira and A.~Zee,
  Phys.\ Lett.\  B {\bf 161}, 136 (1985);
  J.~McDonald,
  Phys.\ Rev.\  D {\bf 50}, 3637 (1994)
  [arXiv:hep-ph/0702143];
  D. E.~Holz and A.~Zee,
  Phys.\ Lett.\  B {\bf 517}, 239 (2001)
  [arXiv:hep-ph/0105284];
  X. G.~He, T.~Li, X.Q.~Li, and H.C.~Tsai,
  Mod.\ Phys.\ Lett.\  A {\bf 22}, 2121 (2007)
  [arXiv:hep-ph/0701156];
  D.~O'Connell, M.~J.~Ramsey-Musolf, M.~B.~Wise,
  Phys.\ Rev.\  {\bf D75}, 037701 (2007).
  [hep-ph/0611014];
   V.~Barger, P.~Langacker, M.~McCaskey, M.J.~Ramsey-Musolf, and G.~Shaughnessy,
  Phys.\ Rev.\  D {\bf 77}, 035005 (2008)
  [arXiv:0706.4311 [hep-ph]];
  V.~Barger, P.~Langacker, M.~McCaskey, M.~Ramsey-Musolf, G.~Shaughnessy,
  Phys.\ Rev.\  {\bf D79}, 015018 (2009).
  [arXiv:0811.0393 [hep-ph]];
   S. M.~Carroll, S.~Mantry, and M.J.~Ramsey-Musolf,
  [arXiv:0902.4461 [hep-ph]];
 X.G. He, T. Li, X.Q. Li, J. Tandean and H.C. Tsai,
 Phys. Lett. B {\bf 688}, 332 (2010)
 [arXiv: hep-ph/0912.4722];
 X.~G.~He, S.~Y.~Ho, J.~Tandean and H.~C.~Tsai,
  Phys.\ Rev.\  D {\bf 82}, 035016 (2010)
  [arXiv:1004.3464 [hep-ph]];
 Y.~Cai, X.~G.~He and B.~Ren,
  arXiv:1102.1522 [hep-ph].



\bibitem{cerilli}
M. Cerilli, N. Fornengo, A. Strumia, Nucl. Phys. {\bf B753}, 178(2006); M. Cerilli, A. Strumia, New. J. Phys., 11, 105005 (2009).



\bibitem{seesaw1}
 P.~Minkowski,
  Phys.\ Lett.\ B {\bf 67}, 421 (1977);
  T.~Yanagida, in {\it Workshop on Unified Theories}, KEK report 79-18 p.95 (1979);
  M.~Gell-Mann, P.~Ramond, R.~Slansky,
  in {\it Supergravity} (North Holland, Amsterdam, 1979)
  eds. P.~van~Nieuwenhuizen, D.~Freedman, p.315;
  S.~L.~Glashow, in {\it 1979 Cargese Summer Institute on Quarks and Leptons} (Plenum Press,
  New York, 1980) eds. M.~Levy, J.-L.~Basdevant, D.~Speiser, J.~Weyers, R.~Gastmans and M.~Jacobs,
  p.687;
  R.~Barbieri, D.~V.~Nanopoulos, G.~Morchio and F.~Strocchi,
  Phys.\ Lett.\ B {\bf 90}, 91 (1980);
  R.~N.~Mohapatra and G.~Senjanovic,
  Phys.\ Rev.\ Lett.\  {\bf 44}, 912 (1980);
  G.~Lazarides, Q.~Shafi and C.~Wetterich,
  Nucl.\ Phys.\  B {\bf 181}, 287 (1981).

\bibitem{seesaw2}
  W.~Konetschny and W.~Kummer,
  Phys.\ Lett.\  B {\bf 70}, 433 (1977);
%
 T.~P.~Cheng and L.~F.~Li,
  Phys.\ Rev.\  D {\bf 22}, 2860 (1980);
%
 G.~Lazarides, Q.~Shafi and C.~Wetterich,
  Nucl.\ Phys.\  B {\bf 181}, 287 (1981);
%
 J.~Schechter and J.~W.~F.~Valle,
  Phys.\ Rev.\  D {\bf 22}, 2227 (1980);
%
 R.~N.~Mohapatra and G.~Senjanovic,
  Phys.\ Rev.\  D {\bf 23}, 165 (1981);

\bibitem{seesaw3}
  R.~Foot, H.~Lew, X.~G.~He and G.~C.~Joshi,
  Z.\ Phys.\  C {\bf 44}, 441 (1989).

  \bibitem{zee}
  A. Zee, Phys. Lett. {\bf B93}, 389(1980).


\bibitem{e-ma}
E. Ma, Phys. Rev. {\bf D73}, 077301(2006).

\bibitem{other-neu}
E.~Ma,
  Mod.\ Phys.\ Lett.\  {\bf A23}, 721-725 (2008);
E.~Ma, D.~Suematsu,
  Mod.\ Phys.\ Lett.\  {\bf A24}, 583-589 (2009);
P.~-H.~Gu, U.~Sarkar,
  Phys.\ Rev.\  {\bf D78}, 073012 (2008);
Q.~-H.~Cao, E.~Ma, G.~Shaughnessy,
  Phys.\ Lett.\  {\bf B673}, 152-155 (2009);
D.~Suematsu, T.~Toma, T.~Yoshida,
  Phys.\ Rev.\  {\bf D79}, 093004 (2009);
P.-F. Perez and M. Wise, Phys. Rev. {\bf D80}, 053006(2009);
 A.~Adulpravitchai, P.~-H.~Gu, M.~Lindner,
  Phys.\ Rev.\  {\bf D82}, 073013 (2010);
T.~Li, W.~Chao,
  Nucl.\ Phys.\  {\bf B843}, 396-412 (2011);
S.~Kanemura, O.~Seto, T.~Shimomura,
[arXiv:1101.5713 [hep-ph]];
 M.~K.~Parida,
 [arXiv:1106.4137 [hep-ph]].



\bibitem{Sher:1988mj}
  M.~Sher,
  Phys.\ Rept.\  {\bf 179}, 273 (1989).

\bibitem{Casas:1994qy}
  J.~A.~Casas, J.~R.~Espinosa and M.~Quiros,
  Phys.\ Lett.\  B {\bf 342}, 171 (1995)
  [arXiv:hep-ph/9409458].

\bibitem{Hambye:1996wb}
  T.~Hambye and K.~Riesselmann,
  Phys.\ Rev.\  D {\bf 55}, 7255 (1997)
  [arXiv:hep-ph/9610272].

\bibitem{Gonderinger:2010yn}
  M.~Gonderinger and M.~J.~Ramsey-Musolf,
  JHEP {\bf 1011}, 045 (2010)
  [arXiv:1006.5063 [hep-ph]].

\bibitem{idm1}
D. Tucker-Smith and N. Weiner, {\it Phys. Rev.} {\bf D 64}, 043502 (2001)[arXiv: hep-ph/0101138]

\bibitem{idm2}
D. Tucker-Smith and N. Weiner, {\it Phys. Proc. Suppl.} {\bf 124}, 197(2003)[arXiv: astro-ph/0208403].

\bibitem{idm3}
D. Tucker-Smith and N. Weiner, {\it Phys. Rev. } {\bf D 72}, 063509(2005)[arXiv: hep-ph/0402065].

\bibitem{idm4}
 S. Chang, G. D. Kribs, D. Tucker-Smith and N. Weiner, arXiv: 0807.2250.

\bibitem {smdm}
T. Hambye, F.-S. Ling, L. Lopez Honorez and J. Rocher, {\it JHEP} {\bf 0907}, 090(2009)[arXiv: hep-ph/0903.4010].

\bibitem{50g}
R.~Barbieri, L.~J.~Hall, V.~S.~Rychkov,
  Phys.\ Rev.\  {\bf D74}, 015007 (2006).
  [hep-ph/0603188];
   L.~Lopez Honorez, E.~Nezri, J.~F.~Oliver, M.~H.~G.~Tytgat,
  JCAP {\bf 0702}, 028 (2007).
  [hep-ph/0612275];
  T.~Hambye, M.~H.~G.~Tytgat,
  Phys.\ Lett.\  {\bf B659}, 651-655 (2008).
  [arXiv:0707.0633 [hep-ph]].


\bibitem{160g}  L.~Lopez Honorez, C.~E.~Yaguna,
  JCAP {\bf 1101}, 002 (2011).
  [arXiv:1011.1411 [hep-ph]].




\bibitem{superCDMS} E. Aprile and L. Baudis (XENON Collaboration) 2Proc. Identification of Dark Matter 2008 Conf.
(Stockholm, Swesen, 18-22 aug. 2008)(Trieste, Italy: Proceedings of Science) pp 1-10 (arXiv:0902.4253[astro-ph.IM);

CDM-II Collaboration2005 eConf C041213(2004( 2529(arXiv:astro-ph/0503583).


\bibitem{tri-bimaximal}
P.F. Harrison, D.H. Perkins, W. G. Scott, Phys. Lett. {\bf B530}, 167(2002); Z.-Z. Xing, Phys. Lett. {\bf B533}, 85(2002); X.-G. He and A. Zee, Phys. Lett. {\bf 560}, 87(2003).

\bibitem{MEG} J. Adam et al., [MEG Collaboration][arXiv:1107:5547].


\bibitem{meg} R. Sawada, (MEG Collaboration), PoS (IHEP 2010) 263.



\bibitem{Ref:Mu2E}
%
J.~P.~Miller [Mu2E collabaration], {\it Proposal to Search for $\mu^- N\to e^- N$ with a Single Event Sensitivity Below $10^{-16}$}.





%
%
%
%
%




\bibitem{Ref:COMET}
%
Y.~Kuno et.al. [COMET collaboration], {\it An Experimental Search for lepton Flavor Violating $\mu - e$ Conversion at Sensitivity of $10^{-16}$ with a Slow-Extracted Bunched Beam}.

\bibitem{Ref:PRISM}
%
Y.~Kuno et.al. [PRISM/PRIME Group], Letter of Intent, {\it An Experimental Search for a $\mu - e$
Conversion at Sensitivity of the Order of $10^{-18}$ with a Highly Intense Muon Source: PRISM}.

\bibitem{four}
N.~Deshpande, T.~Enkhbat, T.~Fukuyama, X.~G.~He, L.~H.~Tsai and K.~Tsumura,
  arXiv:1106.5085 [hep-ph].



\bibitem{kitano}
R.~Kitano, M.~Koike and Y.~Okada,
  Phys.\ Rev.\  D {\bf 66}, 096002 (2002)
  [Erratum-ibid.\  D {\bf 76}, 059902 (2007)],

\bibitem{Aharmim:2009gd}
  B.~Aharmim {\it et al.}  [SNO Collaboration],
  Phys.\ Rev.\  C {\bf 81}, 055504 (2010)
  [arXiv:0910.2984 [nucl-ex]].

\bibitem{Adamson:2008zt}
  P.~Adamson {\it et al.}  [MINOS Collaboration],
  Phys.\ Rev.\ Lett.\  {\bf 101}, 131802 (2008)
  [arXiv:0806.2237 [hep-ex]].

\bibitem{cos-mass} F. De Bernardis, P. Serra, A. Cooray and A. Melchiorri, Phys. Rev. D {\bf 78}, 083535 (2008).

\bibitem{superB} T. Aushev et al., Physics at uper B Factory, arXiv:1002.5012.


\bibitem{Beda:2009kx}
  A.~G.~Beda {\it et al.},
  Phys.\ Part.\ Nucl.\ Lett.\  {\bf 7}, 406 (2010)
  [arXiv:0906.1926 [hep-ex]].
\end{thebibliography}
\end{document}